# Overheat Instability in an Ascending Moist Air Flow as a Mechanism of Hurricane Formation

A.Nechayev, A.Solovyev

Moscow State University, Laboratory of Renewable Energy Sources

## Abstract

The universal instability mechanism in an ascending moist air flow is theoretically proposed and analyzed. Its origin comes to the conflict between two processes: the increasing of pressure forcing applied to the boundary layer and the decelerating of the updraft flow due to air heating. It is shown that the intensification of tropical storm with the redistribution of wind velocities, pressure and temperature can result from the reorganization of the dissipative structure which key parameters are the moist air lifting velocity and the temperature of surrounding atmosphere. This reorganization can lead to formation of hurricane eye and inner ring of convection. A transition of the dissipative structure in a new state can occur when the temperature lapse rate in a zone of air lifting reaches certain critical value. The accordance of observational data with the proposed theoretical description is shown.

## 1. Introduction

An inertial stability of stationary atmospheric vortex offers no difficulties for the theoretical explanation. The existing radial pressure gradient must be balanced by the centrifugal force of rotating air that conserves its whole angular momentum and therefore conserves its rotation axis. This is more of a challenge to explain the origin of a pressure drop. The hurricane pressure deficit is a result of a long time deepening of originally weak tropical depression having the horizontal size ten fold exceeding a troposphere thickness. Attempts to explain the depression growth by means of a certain mechanism of instability face the contradiction: convective instability in the atmosphere (for example formation of a cumulus cloud) allways develops with small time constant and has rather small horizontal dimensions. The latter anyway shouldn't exceed the thickness of troposphere. How a stable huge vortex with a diameter more than 1000 km and with time formation exceeding several days can amplify itself and overcome this fast and small-scale instability was not clear.

 A number of basic theories have been put forward in the mid-sixties of the last century. Later they were developed or rejected. The software of theoretical works has become deeper, but it is difficult to remember a single intelligible and uncontradictory explanation of the physical mechanism of hurricane formation and intensification. The evident reason for it is the exceptional complexity of the phenomenon which incorporate almost all basic spheres of classical physics – hydrodynamics, thermodynamics, phase transitions.

 The first idea which put theoretical researches in motion belongs to Charney and Eliassen (1963). They proposed and analysed the instability mechanism (Conditional Instability of Second Kind – CISK) which basic conception comes to the formula that "cumulus clouds and the large-scale circulation cooperate, rather than compete." Really the tropospheric warming-up is proportional to the vertical flow of water vapor and consequently is proportional to the speed of air lifting. This updraft takes place in areas of low-level air convergence which is created by the pressure drop proportional to tropospheric heating. The mathematical analysis carried out by Charney and Eliassen (1963) proved the possibility of fluctuational instability however the proposed mechanism had a number of serious simplifications and restrictions.

 The theory of CISK has got an approval from some researchers and has been subjected to criticism from others (Ooyama ,1982, 1969; Emanuel, 1989, 1991). The possibility of heat accumulation in the vortex center owing to the moist air convergence in the low levels has corroborated by Ooyama's numerical simulation (Ooyama, 1969).The author of the alternative theory of hurricane Kerry Emanuel (Emanuel,1986, 1989, 1991) was among critics of CISK-theory. His theory was named WISHE (Wind Induced Surface Heat Exchange) and was based on mechanic and thermodynamic conservation laws. Following WISHE-theory the hurricane motive power consists in an increase of an ocean-air heat flux (basically due to the growth of evaporation) at wind strengthening. The presence of deficiencies in basic assumptions of WISHE-theory was specified repeatedly in some works (Smith, Montgomery and Vogl, 2008). Both CISK and WISHE theories pretend for the explanation of the initial phase of hurricane formation. For further intensification other mechanisms are required. In particular the mechanism of reorganization of tropical storm structure during the eye formation remains out of consideration.

Till then a lot of attempts to solve a problem of hurricane formation using purely mathematical methods have been undertaken, however they resulted in conceptual errors or in unproductive discussions. The mathematics deprived of a physical basis is closed on itself. But if the initial physical idea is incorrect or it is insufficiently considered the most effective mathematical methods including numerical simulation turn into the guide who does not know where to lead the blind.

The present work has no purpose to criticize the existing theories of hurricane formation and intensification. An essentially new physical approach considering hurricane as an open nonequilibrium integrated system, having a number of steady states, here is proposed. Between these states transitions are possible. The primary goal of the given work consists in search of the mechanism destabilizing a state of homogeneous atmospheric vortex (cluster, tropical storm) and transforming it into a hurricane vortex with highly redistributed key parameters: wind velocity, pressure and air temperature.

# 1. From cluster to tropical storm. The layer of maximum warming

The hurricane history begins with cluster and tropical depression. Cluster is a gathering of cloudiness where a vertical lifting of moist air is observed within several days. The typical cross-section dimensions of clusters vary from 200 to 1000 kilometers. An atmospheric warming-up in clusters is very weak: no more than 1°C in the top layers however the earth surface has an appreciable pressure deficit (1-2 ГПа). We would like to present here physical circumstance capable to lead to a stable primary pressure drop in the cluster area. Indeed lifting and accumulation of water vapor above condensation level ($z = W$) is accompanied by replacement by this vapor of corresponding volumes of dry air. At heights $z > W$ in a zone of cloudiness air pressure is equal to the sum of partial pressure of dry air and water vapor and it is equal to the pressure (at the same level) of dry air in a zone of clear sky. However hydrostatic pressure of total atmospheric column in the cloudy zone will be less than that in the nearby areas of the "clear" sky because of the replacement of molecules of air by more light molecules of water. The maximum hydrostatic pressure deficit will be equal:

$$\Delta p_h = g \int_W^{H_v} \rho_w \frac{\mu_a - \mu_w}{\mu_w} dz = 0,6 g \int_W^{H_v} \rho_w dz \qquad (2.1)$$

where $\rho_w$ is the density of water vapor; $\mu_a$, $\mu_w$ are molecular air and water weights; $H_v$ is the the maximum height of water vapor lifting; $\rho_w = 1,2 q_s$ where $q_s$ is the saturated mixing ratio.

Assuming that at condensation level ($z$=500m) $q_s$=18 g/kg and that in clusters until 5-6 km heights 80 % humidity remain (Frank,1977), we will receive an evaluation $\Delta p_h$ = 1,6 mb that corresponds to the empirical data (McBride, 1981). The dimensions of cluster (hundreds of kilometers) don't allow to low-level radial flows (which velocity is no more than 1m/s) to liquidate this pressure drop before the new portion of water vapor comes aloft.

In the meteorological literature hurricane is compared to "Carno engine" converting the energy of heat in the mechanical energy (Emanuel, 1991). Extending this analogy it is possible to compare hurricane with an internal combustion engine which can't start without ignition – a starter. In hurricane by such a starter the ascending flow of moist air and the primary pressure drop are. The water vapor serves as "gasoline" which is burnt down in a "fire chamber" of condensation. The hurricane engine starts to work steadily when due to "primary draft" the minimum pressure drop providing an uninterrupted delivery of water vapor to the condensation level is reached.

Hurricane is a huge "tube" providing air transportation across the troposphere from bottom to the top. The rough scenario is that: a pressure drop causes a centripetal flow in a boundary layer, the moist air accelerated by forcing rises to the condensation level. Then "relay race" is picked up by a convection and it lifts the warming-up air to the greatest possible level. Probably the convection receives an aid from lateral pressure gradients "compressing" the central core from all sides. Then jets diverge all around forming characteristic spiral rainbands. The main link in this chain is the hydrostatic pressure drop. It is created by the warmed atmospheric column which is in the center of a vortex. There is a question: why does the warm core of hurricane exists so long time and its temperature anomaly isn't liquidate by horizontal gradients of pressure injecting into the warm center a cold air from periphery? The answer may be obvious. The centrifugal force of the rising and rotating air in the eyewall carries out this protective function! The basic warming-up of troposphere is observed at levels above 2 km. In mature hurricanes the overheat of 7-10 °C can be found at heights of 15 km (Hawkins and Imbembo, 1976)! Moist air in the eyewall rises to 10-13 km keeping its high speed of rotation and providing the vertical and radial advection of angular momentum. The dry air with density $\rho$ does not have buoyancy and can't rise to these heights as the maximum height of its possible hydrostatic updraft can't exceed $\Delta p_h/\rho g$ that for mature hurricanes is no more than 1200m. Hence, pledge of survivability of an atmospheric vortex is the constant lifting of rotating moist air above condensation level.

The pressure in free atmosphere can be divided into two parts: the hydrostatic pressure (pressure of an atmospheric column) and the nonhydrostatic one. Everything that concerns the second category isn't defined by gravity. The nonhydrostatic part of pressure is notable for the big uncertainty. The hydrostatic pressure $p_h$ at some height z is calculated from the integrated equation :

$$p_h(z) = \int_z^H \rho g dz \qquad (2.2)$$

where $H$ is the thickness of the troposphere.

If the air density on the height $z$ decreases by the value of $\Delta\rho(z)$ the total pressure of a column above $z$ decreases accordingly:

$$\Delta p_h(z) = \int_z^H \Delta\rho g dz \qquad (2.3)$$

The change of air density $\Delta\rho$ at the given height is connected with changes of temperature $\Delta T$ and pressure by a relationship:

$$\Delta\rho_z = \rho_z(-\frac{\Delta T}{T} + \frac{\Delta p_z}{p}) \qquad (2.4)$$

Correctness of (2.4) can be estimated from following speculations. The process of tropospheric warming-up in hurricane is isobaric so the correctness of the first term of the right side of (2.4) from here follows. But the density at each $z$ level decreases also because the overlying column of air becomes lighter on the value $\Delta\rho_z$ defined by the formula (2.3). Total change of the surface pressure can be evaluated having integrated (2.4) with the account of (2.3):

$$\Delta p = -g\int_0^H \rho\frac{\Delta T}{T}dz - g^2\int_0^H \frac{1}{p}(\int_z^H \rho\frac{\Delta T}{T}dz)dz \qquad (2.5)$$

It is clear that this bulky formula is the only first iteration in the calculation of the pressure of an atmospheric column in hurricane but nevertheless it gives an evaluation close to a reality. However it is evident that the pressure deficit increases with the increasing of the hight of bubble lifting and with the total overheat $\Delta T$.

The primary warming of the troposphere at hurricane formation occurs in tropical clusters and tropical storms due to latent heat release in the rising moist air parcels. Air receives its high humidity from the ocean surface evaporation and reaches the condensation level due to some poorly explored phenomena, for example, easterly waves or Ekman pumping (Charney and Eliassen, 1963). The tropospheric warming provides the hydrostatic pressure drop between the tropical storm center and its periphery. The certain part of this pressure deficit which can be named by "pressure forcing" (or "forcing" ) always makes free air move.The character of this movement (rectilinear or vortical) depends on the form of surfaces of $p/\rho$ potential ( $p$ and $\rho$ are air pressure and air density). If this surface has a "gap", air streams will direct to its center and a circulation provided by Coriolis force can start around there. If an axisymmetric pressure forcing $\Delta p_h$ with a radius $L$ is applied to the boundary layer with thickness $W$ we must have a centripetal flow here with radial, azimutal and vertical velocities ($u,v,w$ accordingly). The structure of an idealized stationary vortex should satisfy the conditions: $v = u = 0$ at $r = 0,L$ and $w = 0$ at $z = 0$. The presence of wind extremums $v_m$ at $R_m$ (RMW– Radius of Maximum Wind) and $u_m$ at $r_m$ and the existence of updraft area ($w > 0$) in a zone of negative divergence ( divu =u/r+∂u/∂r < 0 ) from here follows. In this area the main warming of the tropospheric column occurs.

The energy of hurricane comes from a temperature difference between ocean and atmosphere. Hurricane is an unclosed system whose energy initially increases and finally dissipates. The heat released from water vapor condensation is transferred to the surrounding air. Surface pressure decreases and the air from all directions is directed toward this place. It seems all should end with it. So it would if an air had no possibility to continue the movement upwards. But the moist air reaching the condensation level has such a possibility. Besides this air coming to the zone of low pressure loses its density which it had on periphery (according to the ideal gas law) and rising upwards it becomes even lighter. Positive feedback is available.

They say that hurricane is an enormous cumulus thunder cloud. But hurricane is not a cloud, it represents the system of clouds participating in a joint rotary motion round the center. Circulation is the main property of an atmospheric vortex. Presence of an axis of rotation gives the birth to special mechanical characteristic – the angular momentum. The law of angular momentum conservation is one of the strongest laws of physics. It is often applied to hurricane air parcels as they rotate around the center. However the angular momentum is constant if vectors of all forces applied to a body or to its part pass through the rotation axis. The forces untwisting hurricane are pressure forces which accelerate air in the bottom part of hurricane in a boundary layer. These forces are not perpendicular to the rotation axis (there exist a so-called inflow angle) that is they have the radial component directed to the axis and tangential component creating linear acceleration. The points on a curve $v (r)$ satisfying $vr = const$ concern not the same air parcel (as should be at the realization of any conservation law) but different parcels and even different streams!

To understand tropospheric warming some simple everyday analogies help. Let's imagine a group of people standing in the middle of an enormous covered stadium and starting upwards the balls filled with warm air. The stadium has the flat roof ( the meteorological analogy is tropopause) and balls rest against it and stop. Infinitely there are a lot of balls, they continuously rise collecting under the roof. Gradually they densely fill the first layer then the second, the third. Those balls that are in the center push aside that over them to the edges – so all the roof is filled. Filling a layer behind a layer balls actually replace the air under the roof by other warmer air. If you imagine as well that they can be slowly blown in the course of lifting they will warm up all atmosphere of the stadium.

In rising parcels ("bubbles") there are two basic confronting processes: adiabatic cooling (-9,8°C on each kilometer of lifting) and an isobaric warming owing to the condensation process according to specific humidity of a parcel $q$. The rate of latent heat release of last process $Q_L$ can be evaluated as:

$$Q_L \sim \frac{L_c}{c_p} \frac{\partial q_s}{\partial T_b} \frac{\partial T_b}{\partial z} w \qquad (2.6)$$

where $T_b$ and $w$ is the bubble temperature and vertical velocity; $L_c$ is the latent heat of condensation ($L_c = 2,5 \cdot 10^6$ J/kg); $C_p$ is the specific heat of air at constant pressure ($C_p= 1000$ J/kgK ).

$$q_s = 0,622 \frac{e_s}{p} \qquad e_s = 6,105 \exp\left[\frac{17,67 t}{t + 243,5}\right] \qquad \begin{array}{c}(0<t<60°C)\\ [e]= mb\end{array} \qquad (2.7)$$

where $e_s$ is the saturated vapor pressure (Bolton, 1980):

The most of this condensation heating goes on the compensation of adiabatic cooling with the lapse rate $\Gamma_d$= 9,8°C/km. Substantially smaller part dissipates in surrounding atmosphere increasing its temperature $T$. Some part remains in a bubble providing its further lifting. Process of warming-up of the atmosphere at the lifting of moist air isn't studied practically. It is possible to assume that the bubble with $T_b > T$ mixes up with environmental air partially. Thus losing an external part of its volume and heat the bubble keeps "the core" with the greatest possible temperature and decreasing in mass moves to the top limit of the buoyancy.

Nobody knows the temperature of a bubble. Modern dropwindsondes don't allow to measure it precisely as their time constants (at 20°C: 2,5s for temperature measuring and 0,1s for humidity (Hock and Franklin, 1999)) and possibilities of spatial resolution give obviously averaged values. There are objective data for functions $q_s(T)$, $T(z)$ and $w(z)$ but concerning the magnitude of $\partial T_b/\partial z$ in (2.6) it is possible to build assumptions only. Generally speaking, in the idealized case of bubble lifting its vertical temperature gradient must satisfy the equation:

$$\frac{\partial T_b}{\partial z} = -\left(\frac{L_c}{c_p}\frac{\partial q_s}{\partial T}\frac{\partial T_b}{\partial z} + \Gamma_d\right) \tag{2.8}$$

Takng into account from (2.2) that in the range 0<t<30°C $\partial q_S/\partial T_b \approx q_S/15$ we can derive for bubble lapse rate:

$$\Gamma_b \equiv -\frac{\partial T_b}{\partial z} = \frac{\Gamma_d}{1+q_s/6} \tag{2.9}$$

At rather high altitudes where $q_s$ is small enough $\partial T_b/\partial z$ should be close to dry adiabatic lapse rate. In the bottom layers $\partial T_b/\partial z$ should be between its minimum (the bubble keeps temperature) and the lapse rate in the surrounding atmosphere. Anyway the rate of latent heat release $Q_L$ is proportional to $\partial q_S/\partial T_b \cdot \partial T_b/\partial z \cdot w(z)$ and therefore should have a layer or a zone of maximum values as $\partial q_S/\partial T_b$ decreases with height and $\Gamma_b$ and $w(z)$ increase (Fig 1). Really the magnitude of $\partial q_S/\partial T$ at levels from 1 to 4 km decreases by a factor of 2-3, and the magnitude of $w$ increases in clusters at the same heights by a factor of 3-4 (McBride, 1981). So there should be a layer $z_m$ where the warming goes faster than in other layers (Fig. 1,2). It leads to the reduction of the lapse rate $\Upsilon$ in underlaying layers, as $\Upsilon \approx -(T_m - T_W)/(z_m - W)$ where $T_m$ is the air temperature in a layer of the maximum warming-up, $T_W$ is the temperature at condensation level which practically doesn't change.

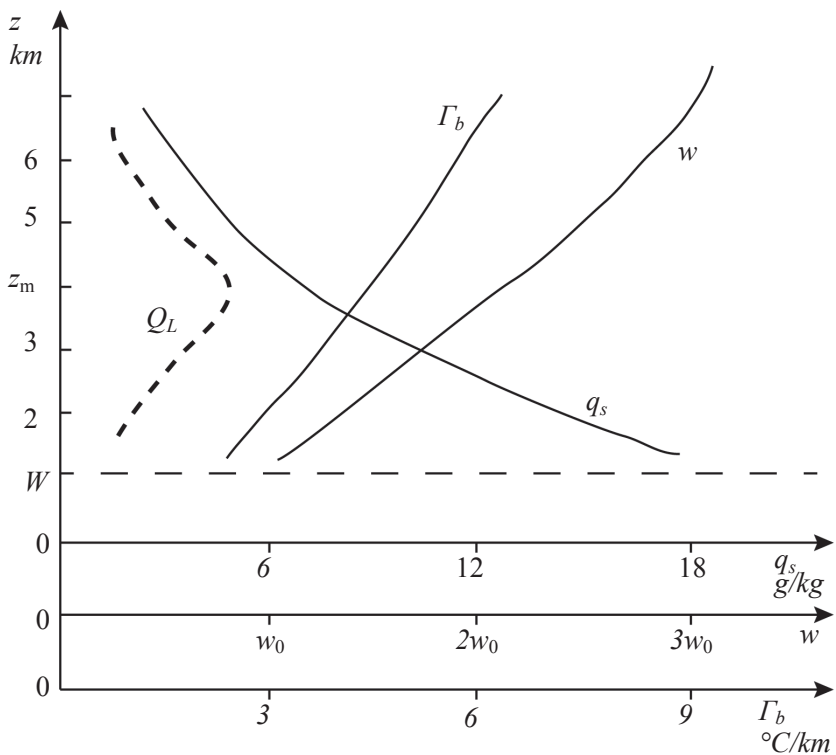

Fig.1 *Qualitative altitude dependences of $q_S$, $w$, $\Gamma_b$ and their product $Q_L$ – the rate of latent heat release on the level of maximum warming-up $z_m$; W is the condensation level.*

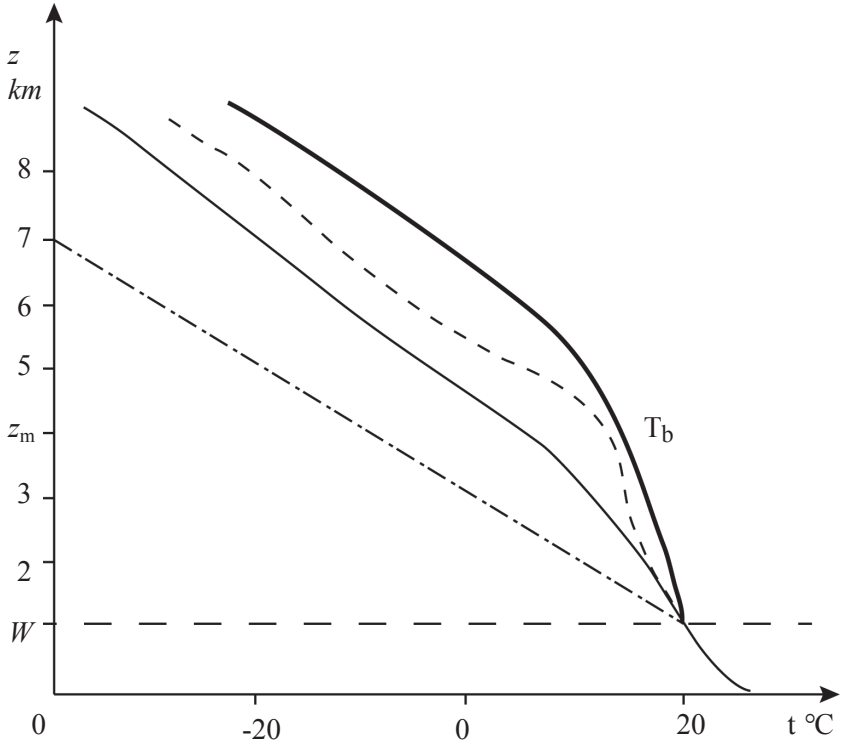

Fig. 2 Qualitative temperature soundings illustrating the heating of the atmosphere by the ascending moist air. The temperature of "bubble"(heavy line), the temperature of atmosphere before (thin line) and after (dashed) the heating. Dash-dotted line is the dry adiabat. $z_m$ is the level of maximum warming-up

The warming of the tropospheric column in the center of the vortex provides the horizontal deficit of hydrostatic pressure which is always maximum near the surface. The magnitude of pressure gradient $\partial p/\partial r$ changes weakly within the boundary layer $(0 < z < W)$ as the main warming is characteristic of the top and middle layers of troposphere. So the integral form of the averaged mass conservation law gives:

$$w_0 \pi r_m^2 = -2\pi r_m W u_m \qquad (2.10)$$

where $u_m$ is the maximum radial velocity (the centripetal direction of flow is negative); $w_0$ is the vertical velocity at $z = W$; $r_m$ is the radius of maximum radial wind (we take it approximately equal to the radius of updraft area which exactly is defined by the condition: div$u < 0$). Certainly $u_m$, $w_0$ and $\rho$ are averaged accordingly within $r_m$ and $W$. We have:

$$w_0 = -\frac{2u_m}{r_m} W \qquad (2.11)$$

Dependences $v(r)$ and $u(r)$ give the information of velocity field but don't reflect completely the process of centripetal movement. Points on these curves correspond as a rule to different streamlines and some sections belong to substantially different zones. For example, the area $r < R_m$ in some vortex represents a zone of "solid-body" rotation ($v = \omega r$, $\omega = v_m/R_m$) in which we can use the equation of forces balance but the Bernulli law is incorrect. In a zone $r > r_m$ (within a separate streamline) it is good to use both forces balance and Bernulli law which gives for $r = r_m$ at constant heights and small $w$:

$$\Delta p = p_0 - p = \frac{1}{2}\rho(v_{um}^2 + u_m^2) \qquad (2.12)$$

where $p_0$ is the pressure in some reference point $r = L$; $v_{um}$ is the azimutal velocity at $r = r_m$

For steady state conditions it is obligatory that stream lines of an atmospheric vortex had closed trajectories that is the horizontal flows must turn to vertical direction. The moist air reaching the condensation level can continue its lifting only in the form of buyoant warm bubbles. It occurs if the bubble is surrounded by sufficient volume of the air with $T < T_b$. In other words, if the flow $I$ approaches condensation level only some part equal $I/k$ where $k > 1$ can continue the upward motion. Actually $1/k$ is a part of the full area occupied by bubbles. To understand better the structure of an atmospheric vortex let's make a mental experiment. Let's assume that in the layer $0 < z < W$ there is a horizontal pressure deficit $\Delta p_h$ creating a centripetal air inflow. The maximum possible magnitude of this flow is: $I_{max} = 2\pi r_m \rho |u_m| W$, and the maximum energy of each parcel according to the Bernulli law *(v=0)* must be equal $\rho u_m^2/2 = \Delta p_h$. This horizontal flow turns into the vertical one which for the reason specified above can't exceed $I_{max}/k$. To conserve the inflow mass and its energy ($\Delta p_h$=const) it is possible only having reduced $u_m$ in $k$ time and having established circulation with the speed $v_{um}$ satisfying the energy condition:

$$v_{um}^2 + \left(\frac{u_m}{k}\right)^2 = u_m^2 \qquad (2.13)$$

As a result we come to the relationship

$$v = u\sqrt{k^2 - 1} \qquad (2.14)$$

which doesn't depend neither on dimensions of a vortex nor on pressure drop magnitude and is defined only by physics of the process of convection at the condensation level (factor $k$). There are numerous observations of hurricanes with a surprising constancy of the inflow angle $\alpha$ ( $tg\alpha = u/v$) equal 22-25° in a wide area $r > r_m$ at low levels (Frank, 1977). As $tg22° = 0,4$ so $k = 2,7$.

The primary pressure drop should cause centripetal air flow which under corresponding conditions (Coriolis force or a certain vortex-germ) can lead to a stationary circulation. According to cluster observations (McBride, 1981; McBride and Zehr, 1981) circulation especially in a boundary layer is the absolutely necessary condition for hurricane formation. Only it due to radial flow convergence is capable to support an air ascent in the vortex center and to provide here the maximum warming-up and the minimum heat dissipation as azimuthal speeds are minimum. Just the circulation provides the positive feedback $w \rightarrow \Delta T \rightarrow \Delta p_h \rightarrow w$ due to increasing of the maximum radial velociy $u_m$. The moist air ascent leads firstly to a warming-up of top layers, then the middle and bottom layers become heated that increases the horizontal pressure drop $\Delta p = p_0 - p$ which is applied to all boundary layer $0 < z < W$. For considered air lifting area $r < r_m$ there exists a monotonously increasing function $w_0(\Delta p_h)$ where $w_0$ is a vertical velocity at condensation level. According to (2.11), (2.12), (2.14) $w_0$ may be expressed by the formula:

$$w_0 \approx \frac{2(2\Delta p_h)^{1/2}}{\rho k r_m} W \qquad (2.15)$$

Formula (2.15) reflects the essence of initial (stable) intensification of tropical storm.

In the presence of favorable external conditions (the main condition is the absence of vertical wind shear) necessity for the primary air lifting disappears. A tropical storm can support itself and can leave his "cradle" and set out on a big voyage.

# 3. From tropical storm to hurricane. Dissipative structure. "Flow-forcing" characteristic.

The fundamental basis of hurricane formation is probably the conflict between two processes: the increasing of pressure forcing applied to the boundary layer and the decelerating of the updraft flow due to air heating. The leading role is played by the temperature field especially in the layer of maximum warming. In this area we have a classical dissipative structure with two key parameters: the updraft velocity of moist air and the temperature of surrounding atmosphere (Nechayev, 2011) . This dissipative structure is controlled by three equations which can be written in the simplified form. The first is the equation of heat balance for an area $r<L, z>W$.

$$\frac{\partial T}{\partial t} + \frac{\partial T}{\partial r} u = \frac{T_b - T}{\tau_T} + \frac{L_c}{c_p} \frac{\partial q_s}{\partial T_b} \frac{\partial T_b}{\partial z} w\Phi(v^2) + \nu_T \nabla^2 T \qquad (3.1)$$

The first term on the right-hand side of equation (3.6) is responsible for heat exchange of a bubble with the surrounding atmosphere going with the time constant $\tau_T$ (it is maximum in the top layers of troposphere and it is minimum in bottom, where $T_b \sim T$); the second term – a rate of latent heat of condensation (it is minimum in the top layers and maximum at level $z_m$); $\Phi$ is a heat transfer effectiveness ratio ($\Phi <1$) depending on geometrical parameters of a bubble and on degree of updraft turbulence ( actually on azimutal velocity). $v_T$ is a turbulent heat factor depending on velocity field. Boundary conditions correspond to axial symmetry : $\partial T/\partial r = 0$ ($r=0, L$). Vertical velocity of bubble $w$ should satisfy the equation of motion in which the hydrostatic part is excluded and a standard Newtonian drag force is added:

$$\frac{\partial w}{\partial t} + \frac{\partial w}{\partial z} w = \frac{T_b - T}{T} g - C_b d_d^2 w^2 + v_w \nabla^2 w \qquad (3.2)$$

where $d_b$ is the diameter of a bubble, $C_b$ is a proportionality factor; $v_w$ is a turbulent viscosity factor.

The third equation determines the quantity of a total air flow $I$ through the vortex:

$$\int_0^L 2\pi r \rho w dr = I(t) \qquad (3.3)$$

The redistribution of vertical air flow is controlled by the equation (3.3) according to the vertically averaged in the boundary layer mass conservation law:

$$\frac{w_0}{W} = -(\frac{\partial u}{\partial r} + \frac{u}{r}) \qquad (3.4)$$

where $w_0$ serves as the initial and boundary condition of the equation (3.2) at z = W.

At the initial stage the total air flow $I$ grows with the increasing of $\Delta p_h$ as the vertical velocity $w_0$ increases according to (2.15). At some instant the reduction of the lapse rate in a layer of the maximum warming-up will decrease the bubble acceleration. Therefore $w_m$ will increase more slowly then will cease to increase and even can start to decrease while $T_b \to T$ (Fig. 3). However $T_m$ and pressure forcing $\Delta p_h$ will increase as the heating of total tropospheric column continues $(w > 0)$. There the conflict known in the physics (basically in semiconductor electronics) between an air flow (current) and pressure drop (voltage) takes place. In electronics the presence of a "falling" segment of the current-voltage characteristic leads to the classical instability (for example the well-known effect of Gunn). Let's introduce the so-called "flow-forcing" characteristic $I(\Delta p_h)$ as a certain analogue of current-voltage characteristic. In our case there should be a falling segment (a N-shape segment) of characteristics $w_m(T_m)$ and $I(\Delta p_h)$ where $\partial w/\partial T <0$ and $\partial I/\partial p<0$ that gives a fundamental basis for redistribution of $w$ and $T$ leading the system to a transition to the new state (Fig 3,4).

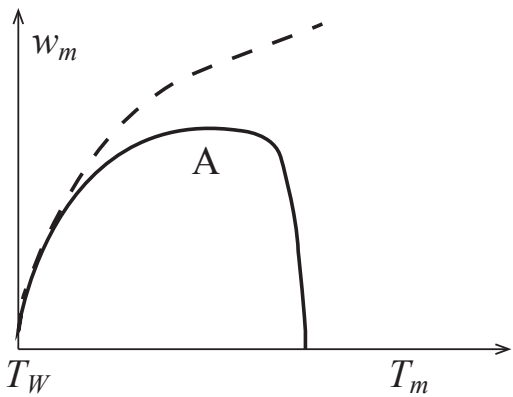
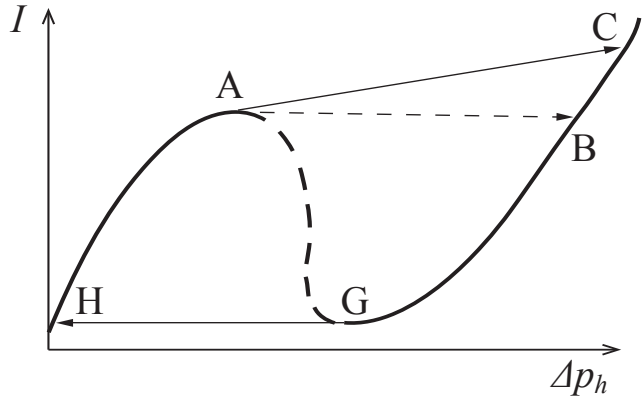

Fig. 3 *Qualitative form of the dependence $w_m(T_m)$ in case of critical overheat (solid line) and without air flow decelerating (dashed line).*

Fig. 4 *The flow-forcing characteristic $I(\Delta p_h)$ of supposed dissipative structure. The transition ABC is accompagnied by the redistribution of $w(r)$ and the warming-up of a central column*

We can suppose each standard cumulus cloud (and all the more a "hot tower") to be an independent dissipative structure in which a lifting of moist air and an accompanying warming-up of troposphere occurs. Whether it will come to a critical condition depends on many parameters including the geometrical parameters of structure: its cross-section dimensions, the height of air lifting and the moisture contents. If certain forcing $\delta p_h$ is present in the low-levels ($\delta p_h = l_c \partial p_h/\partial r$, where $l_c$ is the radial size of cloud base) the dissipative structure of the cumulus cloud must have an initial "growing" segment of the characteristic $i(\delta p_h)$ ). The overheat of some layers can result in decreasing of total $i$ (Fig.5b ). The growth of $\delta p_h$ (for example, in the zone of nonhydrostatic pressure gradient) may overcome the critical stage (Fig.5c) and result in chaotic streams redistributions.

Cumulus clouds of a hurricane providing the basic vertical air flow I are summarized in a total dissipative structure. The characteristic $I (\Delta p_h)$ of hurricane is the sum of "small" characteristics $i (\delta p_h)$ of separate clouds. Some of them can already have the local N-shapeness of $i(\delta p_h)$, others still "grow" upwards (Fig. 5). When the large aggregate of clouds-cells will start to feel the weakening of an ascending flow (due to reduction of $|\partial T/\partial z|$), the total state of a tropical storm will approach the critical point: $\partial I/\partial p \rightarrow 0$. With the beginning of the reorganization the clouds in the central zone (future eye area) become "exhausted" (see observations of hurricane Gladys (Gentry and Fujita, 1970)) and the clouds on periphery (eyewall area where $\Upsilon$ is more) will be activated (amplified) due to local increasing of nonhydrostatic pressure.

If there is no large-scale circulation which accumulates a tropospheric warming-up in the central zone the clouds-cells which are in this zone won't have a chance to reach the necessary "collective" N-shapeness leading to the hurricane formation.

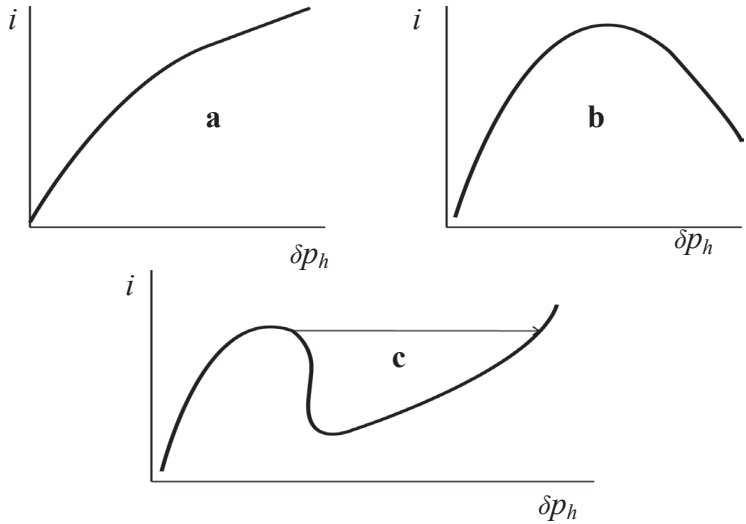

Fig. 5  Three types of cumulus cloud with different dissipative structures: a – normally growing cloud without overheat (zone of maximum lapse rate); b – "exhausted" cloud with overheat and stable decrease of updraft (hurricane eye); c – cloud subjected to the flow redistribution (eyewall,"hot tower"). $\delta p_h = \partial p/\partial r \cdot l_c$ where $l_c$ – radial size of cloud

# 4. Evaluation of critical conditions. Accordance with empirical data.

Certainly, the analysis of critical conditions for such dissipative structure is impossible without numerical simulation. We will try to evaluate these conditions proceeding from simple reasons. Two conditions are necessary for simultaneous lifting and condensation of moist air: $T_b > T$ and $\partial T_b/\partial z < 0$. The time constant of relaxation of liquid phase in clouds at $0°C < t < 20°C$ makes approximately 1s (Korolev and Isaac, 2006). At small $w$ (in clusters $w \approx 10^{-2}$ m/s) the process of condensation apparently must correlate to the lifting velocity (hardly the bubble moves jerky). Researchers believe (Shea and Gray, 1973) that at the beginning of updraft the bubble has a very small overheat (an order of 1°C) which increases in process of lifting in cold layers of atmosphere. Bubble decelerating should come when its minimum lapse rate, defined by the formula (2.9), becomes equal to the lapse rate of an atmosphere. Hence, the critical state of the dissipative structure may require the condition:

$$\Upsilon_{cr} = \frac{\Gamma_d}{1 + q_S/6} \qquad (4.1)$$

where $q_S$ is the saturated mixing ratio of a bubble at the given height.

For the supposed level of the maximum warming-up (650 mb is the level of the typical hurricane inversion) $T = 5\ °C$, $q_S \approx 10 g/kg$. According to the formula (4.1) $Y_{cr} = 3,8\ °C/km$. We will compare this result to the data of observations published in Frank, 1977. Table 1 gives the average temperature soundings for standard tropical atmosphere, clusters and weak hurricanes.

TABLE 1

| Height | Height | Jordan | Cluster | Eye | RMW | 660 km |
|---|---|---|---|---|---|---|
| mb | km | t °C | t °C | t °C | t °C | t °C |
| sfc | 0 | 26,3 | 26,1 | – | 24,5 | 25,7 |
| 900 | 1,0 | 19,8 | 20,7 | 22,1 | 21,1 | 20,2 |
| 800 | 2,1 | 14,6 | 15,4 | 18,7 | 16,7 | 15,5 |
| 700 | 3,3 | 8,6 | 9,4 | 14,3 | 11,3 | 10,2 |
| 600 | 4,7 | 1,4 | 2,0 | 10,0 | 5,0 | 3,3 |
| 500 | 6,3 | -6,9 | -5,7 | 2,7 | -2,3 | -4,7 |
| 400 | 8,3 | -17,7 | -15,8 | -7,2 | -11,2 | -14,9 |

The minimum lapse rate ($Y = 3\ °C/km$) can be seen in a hurricane eye in the layer between 600 and 700mb where the typical inversion of mature hurricanes is observed (Hawkins and Imbembo, 1976; Franklin, Lord and Marks, 1988). Apparently this level corresponds to the layer of maximum warming-up. In eyewall area the lapse rate is more than 4 °C/ km. Probably the magnitude of 3°C/ km have exceeded the critical gradient of 3,8 °C/ km as the dissipative structure was already reorganized.

The maximum temperature of a bubble is defined by the equation (2.8) if the boundary condition $T_b(z=W)$ and humidity $q$ are known. If $q=18g/kg$ the bubble temperature at the height of 3,7 km (650mb) will be equal to 11 °C. The eye temperature of a weak hurricane at this height (Table 1) is close to 12 °C; the typical initial temperature of the 650mb-inversion (Hawkins and Imbembo, 1976; Franklin, Lord and Marks, 1988) is equal to 13 °C. Other words, the reorganization of the dissipative structure probably begins at the equality of a bubble and surrounding atmosphere lapse rate and close magnitudes of their temperatures. Unlike the bubble lifting to the level of free convection there is no full blocking of motion in the layer of maximum warming, an ascending flow remains but the reorganization of its structure occurs.

It is possible to use the concept of critical overheat $\Delta T_{cr}$ connected with the critical lapse rate by the relationship:

$$\Delta T_{cr} = T_W - Y_{cr}(z_m - W) - T_0 \qquad (4.2)$$

For the supposed layer of maximum warming (650 mb) this critical overheat is approximately equal to 5 or 6 °C. So we can evaluate the time of attainment of critical state. According to (2.6) and (3.1) this time is inversly proportional to $Q_L \Phi$. Taking for clusters $w=10^{-2} m/s$, $q_s=10g/kg$, $\partial T_b/\partial z = -4\ °C/km$, we obtain for $Q_L$ the magnitude $1,3 \cdot 10^{-4}\ °C/s$ which gives (for $\Delta T_{cr}=5\ °C$) roughly 12 hours. As far as the characteristic time of hurricane formation averages 5 days we can conclude that the unknown factor $\Phi$ must be equal to $10^{-1}$.

In the area $r < r_m$ the lifting of moist air results in the tropospheric warming and the pressure deficit ($p_0 - p_h$) increases accordingly. The initial distributions of $w(r), T(r)$ within this area are close to homogeneous. The average temperature grows slowly in all the layers so the pressure deficit increases, vertical speed $w_0$ grows also. Streamlines for this stage are represented in Fig 6a. As the center of the structure always has a weak overheat the lapse rate in the layer of maximum warming-up reaches its critical magnitude near $r = 0$. The dependence $w(T)$ in a critical zone can be sharp enough: vertical velocity can reduce to zero at changes of $T$ less than one degree. The reduction of $w_0$ near the center (due to decelerating of air lifting in the layer of the maximum warming-up) leads to corresponding increase of $|divu|$ (due to narrowing of the zone of lifting) and to increase of $w_0$ in areas remote from the center. The redistribution of initial vertical speed at condensation level $w_0$ occurs: in the center it falls to a minimum, on periphery considerably increases (the total vertical flow is constant). The value of $w_0$ serves as a boundary condition at $z=W$ for the equation (3.2). To provide the air updraft in a new place pressure forces must be redistributed too: according to N-shape section of the characteristic $I(\Delta p)$ the domain of nonhydrostatic pressure forces must be formed. Really such domain exists in hurricanes in the zone of sharp pressure gradient within the area of eyewall. The centripetal forcing of rotating jets should create the zone of ring-shaped domain adjoining to the RMW from the inside. This domain (area of strong pressure forces) decelerates and reorients centripetal flows and also provides the basic updraft in the zone of inner ring of convection. Really the eye wall which has a width of 1-5 km always adjoins RMW more subcentral only (Jorgensen, 1984). Reorganization process can be avalanche. Decelerating of radial flows near the center increases the local nonhydrostatic pressure, air flows will search for a new way upward. The flows with $u>0$ will dissipate temperature (the second term in the left side of the equation (3.1)), facilitating air lifting. In addition rising and rotating jets dissipate its own heat much less (in the equation (3.1) factor $\Phi$ reduces ). Thus on some distance from the center there will be created the zone of rotation updraft (an inner ring of convection) where the temperature gradient will be closer to the standart lapse rate of tropical atmosphere (Fig. 6b).

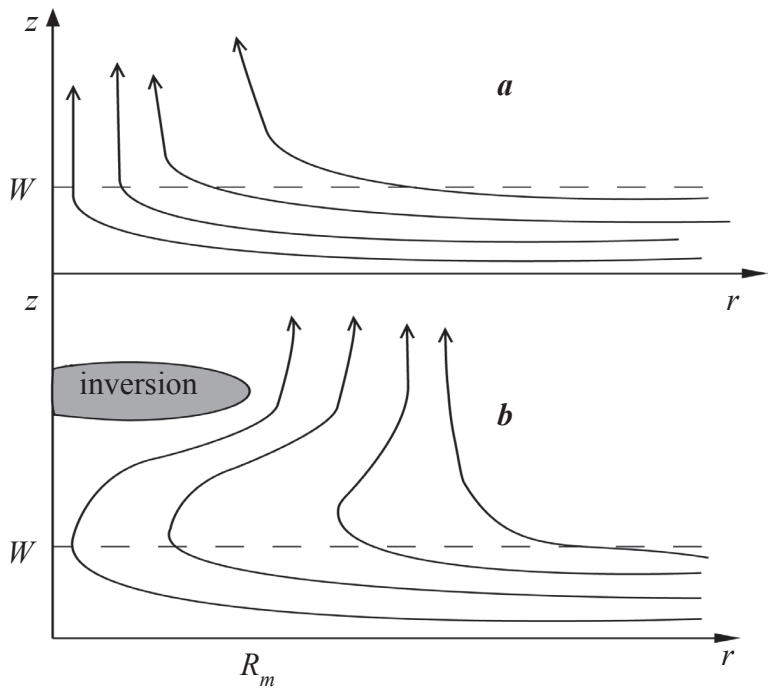

*Fig. 6 Schematic depiction of secondary circulation of a tropical storm (a) and hurricane (b). The ascending streamlines rounding the inversion form the outflow zone where divu>o and an air descent may occur*

Primary inhomogeneity in a tropical storm incurs a role of finite amplitude fluctuation. It just has a mode of an order when the center is heated up. Sharp redistribution begins when $w(T)$ enter into some critical area which corresponds to point A on characteristics $w(T)$ and $I(\Delta p_h)$ (Fig 3,4). Stratification begins: in the vortex center the vertical velocity falls, the temperature continues to grow. In the zone of periferal updraft $w$ increases as the lapse rate here increases. The nonhydrostatic forcing increases too. On the external "flow-forcing" characteristic $I(\Delta p_h)$ the transition A → B to the state with non-uniform distributions $w(r)$, $T(r)$ occurs.

The appearance of strongly pronounced inversion in a hurricane eye at level 650 mb can hardly be explained by adiabatic heating of air due to its descent. Addressing to the mechanism of overheat instability described above it is possible to explain the given "hump" as the further increase of inversion at a layer of the maximum warming-up which has blocked the air lifting and led to sharp redistribution of ascending flows. Its lateral part was destroyed by turbulent dissipation of rotating jets but the central part has remained as the eye rotation goes with constant angular momentum and is "laminar". It is possible to assume that 650 mb is exactly the level where according to (2.6) maximum warming-up occurs.

When the criterion (4.1) is satisfied and the redistribution of vertical flows starts the warming of the center ( especially in the top-levels) doesn't stop as $w>0$. The pressure continues to fall. This process can amplify itself as the rotating eyewall starts to play the role of a "chimney" isolating a warm core from cold surrounding atmosphere. In terms "flow - forcing" a critical condition of the storm-hurricane transition comes when the full lateral flow of air passing through hurricane boundary layer ceases to grow when the pressure drop (forcing) increases. On the characteristic $I(\Delta p_h)$ it is a point A – the beginning of N-shape segment (Fig.4). Transition A→B corresponds to warming of hurricane center with formation of an inner ring of a convection – the area of high azimuthal and vertical speeds. The further pressure drop is accompanied by the contraction of the inner ring. Probably it occurs because the rotating jets pushed to the center by pressure gradient disseminate the inversion as though "eating" its edges.

Secondary circulation (radial and vertical air flows) is regulated basically by the mass conservation law. Scalar fields of temperature and pressure form the original "skeleton" where the velocity vectors try to achive the closed streamlines. Inversion forming in the hurricane center (Fig. 6b) rejects flows going upwards out of the center creating at heights above kilometer a zone of outflow with $u > 0$ and div $u > 0$ that probably causes air descent and warming. The similar idea was stated by Shea and Grey, 1973.

The centripetal flows in the boundary layer carry away the water vapor to the inner ring of convection with the average radial velocity. A radial wind overcomes the distance of 20 km less than for 1 hour. The lack of water vapor is fatal for convection, the warm core must be cooled. The main hurricane force comes from the inexhaustible source of water vapor. The inner ring of convection collects nearly all the water vapor arriving from periphery. It is possible to prove that the evaporation from an ocean surface can't restore the loss of water vapor. Really, maximum speed of evaporation in a tropical zone of the World ocean doesn't exceed 3000 mm a year. It makes about 9 kg of water a day (on the 1 m² of surface area). The corresponding boundary layer column (with the thickness W) contains the quantity of the water vapor equal to $1,2qW$. For $q = 18$g/kg and $W=500$m it is about 10 kg. According to average radial speed (3-5 m/s) all water vapor will be removed from the central zone in some hours, and the former quantity will be restored only in a day.

After hurricane landfall the unlimited source of water vapor (warm ocean) disappears. Warming-up of a tropospheric column on all its height decreases. On the "flow-forcing" characteristic $I(\Delta p_h)$ it corresponds to the movement along the segment B → G (Fig.4). The states on this branch correspond probably to extratropical cyclone which differs both from a tropical storm (segment OA) and from mature hurricane. To get the state of extratropical cyclone it is possible only coming back along a branch BG thanks to a hysteresis of the characteristic $I(\Delta p_h)$.

The proposed mechanism of overheat instability is quite universal and can work not only in hurricane but also in other atmospheric vortex which have an ascending moist air flow in their center. We will consider below some observational data of polar lows (Businger, 1987) and try to explain them using the concept of critical overheat.

Polar low as a natural phenomenon has been described for the first time in the mid-eighties of the XX-th century. Its surprising structural similarity to a tropical cyclone and its main difference – smaller spatial scale (<500 km) and essentially smaller time of generating (within one day) – all indicates a resemblance of physical mechanisms of formation. Polar lows arise during winter time in high latitudes. For their formation some conditions are necessary: first of all is the presence of a wide area of open water and of a neighboring big air mass with a low temperature (may be an ice shield or a continental land). Polar low generation is favored by the layer of cold air extending over an open water surface and by the appearence of the 500mb-trough.

As water temperature can't decrease below 4°C the surface air is appreciably overheated regarding to the surrounding atmosphere and its elevated humidity creates preconditions for a warming-up of an atmospheric column under condition of air lifting above the condensation level. The pressure drop from 500mb-trough is transferred to the surface in the form of insignificant depression (Businger, 1987) which has started to play the role of "primary draft". It provided the lifting of moist air with vertical velocity about $10^{-1}$m/s (Businger, 1987, Emanuel and Rotunno, 1989) and primary circulation in the bottom layers.

Strong evaporation owing to the temperature difference between the open water and surrounding air gives a quite good source of water vapor. However the low moisture content ($q = 7$-$8$ g/kg) and insufficiently large area of open water (in comparison with tropical cyclones) limits life time of the polar low by approximately 12 hours.

The dissipative structure of polar low should be reorganized according to the formula (4.1) when the vertical temperature gradient in a layer of the maximum warming-up reaches its critical value.

Let's consider theoretically the process of polar low formation using the observational data of (Businger, 1987) presented in Fig. 7 (the scale is changed for clearness). Qualitative curves show the temperature soundings obtained with the interval of 12 hours in the zone of formation of polar low over the Bering sea. The first sounding reports about initially stable atmosphere: the temperature drop on the first 2,5 km makes more than 5°

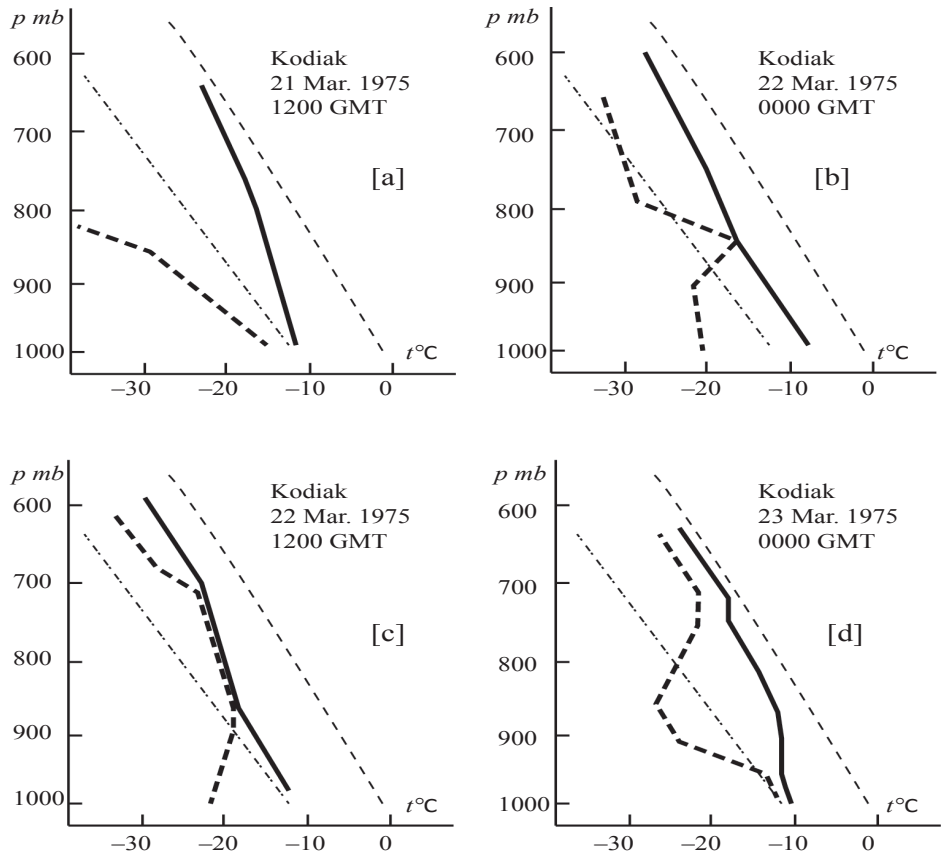

*Fig.14 Temperature soundings taken in the time of polar low formation. Temperature and dew-point profiles ar eindicated by the heavy solid and dashed lines. Dry adiabats are shown as dash-dot lines, moist adiabat as thin dashed lines. Data are taken from (Businger, 1987)*

C/km. In 12 hours after the intrusion of a cold air the bottom part of troposphere was cooled on the average of 5°C that resulted in lapse rate increase to a dry adiabatic. There was the convection first sign – a condensation at the 850mb level (Fig 7b ) that prove the updraft of the air with low mixing ratio. In 12 hours the full influence of 500mb-trough became obvious: a strong updraft with the condensation in the layer about 2 km thickness occurs with an additional cooling of the middle layers due to the same trough. The atmosphere warming-up due to latent heat release and heat exchange between the superheated parcels of moist air and the atmosphere has started in a convection-condensation zone. However at this stage only insignificant general heating of all layers (till 500mb level) compared with the previous sounding can be distinguished (Fig. 7c). Following sounding was made, apparently, when the reorganization of dissipative structure already was finished (Fig.7d ). At 850mb level (probably it was the level of the maximum warming-up $z_m$) the air temperature has grown by 6°C (Businger, 1987) what resulted in the lapse rate decrease in the underlaying layers to 4°C/km that could be a critical magnitude for this case (just as in case of tropical cyclones) . The surface bubble lapse rate according to (4.1) was 3,8° C/ km. But critical stage becomes when the bubble and the surrounding air reach the identical magnitudes of its temperature and lapse rate at the same time. Near the surface the bubble had a rather overheat ($T_b = 4°$ C, T = -10° C) and the rate of latent heat release was small due to practical absence of condensation. That is why the 850 mb level could be the critical one and the maximum warming-up was observed here (Fig.7d). Though the magnitude of $q_s$ at this level was less than the surface one the vertical velocity and the lapse rate of a bubble might be higher and the rate of heating here might have a maximum.

Supposing for this layer (850 mb) $w=10^{-1}$ m/s, $q_s=5$ g/kg, $\partial T_b/\partial z = -6$ °C/km, we obtain for $Q_L$ the magnitude $5 \cdot 10^{-4}$ °C/s which gives for average time of polar low formation ($\Delta T_{cr}=5$ °C, $\Phi =10$ ) 25 hours which corresponds to the observational data ( Businger, 1987, Fig.7)

# 5. Conclusions

Results of the given work allow to propose the following logic chain for the description of hurricane formation.

**1.** There are some extensive areas over a sea surface where a lifting of moist air occurs for a long time.

**2.** Due to the condensation whithin moist air parcels a latent heat release warms-up surrounding atmosphere. The rate and the magnitude of this warming is proportional to the velocity of air lifting and to specific humidity of surface air. This warming is rather weak (some degrees of Celsius) and dissipates if only there is no circulation in a zone of air lifting. Cyclonic circulation assosiated with the convergence in bottom layers cause stable vertical lifting of moist air and heating in the center core of the vortex.

**3.** The rate of tropospheric warming has a maximum in some layer which can reach some kilometers of thickness. The warming of tropospheric column occurs on all height of moist air lifting and it increases pressure deficit between the center of a vortex and its periphery. The more the height of air lifting and the less the vertical wind shear, the deeper the pressure fall. Simultaneously with the general warming of central column there is an accelerated warming in some layer where the lapse rate decreases and hamper the convective lifting.

**4.** The increasing of air inflow in the vortex center due to growth of pressure drop conflicts to the decelerating of an ascending flow in a layer of maximum warming-up. As a result the structure of tropical storm reorganizes forming an eye and an eye wall – a ring-shaped zone of lifting. This convective ring starts to play a role of chimney effectively protecting a warm core of hurricane from the penetration of cold streams from the periphery. As a result a heat dissipation in the hurricane center is extremely low.

**5.** The hurricane which possess the huge inertia momentum keeps it with a rotation axis due to radial pressure gradients which are dynamically compensated by rising and rotating moist air. The quantity of this air is restored not by means of sea evaporation (it goes too slowly) but due to centripetal winds from the periphery.

**6.** Transformation of tropical storm into hurricane (eye formation) occurs presumably while the stable ascending flow in the vortex center starts to feel the decelerating due to tropospheric overheat. It can occur at some critical temperature lapse rate. The rough estimate of this lapse rate gives the value 3,8 °C/km.